\documentclass[aps,prb,floatfix,reprint,twocolumn,superscriptaddress,citeautoscript]{revtex4}
\usepackage[latin1]{inputenc}
\usepackage{dcolumn}
\usepackage{graphicx}
\usepackage{fixmath}
\usepackage[colorlinks=true,linkcolor=black,%
citecolor=black,urlcolor=blue]{hyperref}

\usepackage{amssymb}
\usepackage{amsmath}
\usepackage{amsthm}
\usepackage{bbm}
\usepackage{bm}
\usepackage{multirow}
\usepackage{psfrag}
\usepackage{mathrsfs}
\usepackage{footmisc}

\newcommand{\fett}{\mathbf}
\newcommand{\ve}{\bm}

\newcommand{\te}[1]{\fett{\underline{#1}}}

\newcommand{\etal}{\emph{et al.}}
\DeclareMathOperator{\erfc}{erfc}
\DeclareMathOperator{\erf}{erf}

\newcommand{\itap}{Institut f\"ur Theoretische und Angewandte Physik (ITAP),%
  Universit\"at Stuttgart, Pfaffenwaldring 57, 70550 Stuttgart, Germany}

\newcommand{\udem}{Département de Physique, Université de Montréal,
Montréal, Québec, Canada H3C~3J7}

\newcolumntype{d}{D{.}{.}{3}}

\begin{document}
\title[Wolf summed polarizable silica force field]{Direct Wolf summation of a polarizable force field for silica}

\date{\today}

\author{Peter Brommer}
\altaffiliation[Current Address: ]{\udem}
\email{peter.brommer@itap.uni-stuttgart.de}
\author{Philipp Beck}
\author{Andreas Chatzopoulos}
\affiliation{\itap}
\author{Franz G\"ahler}
\affiliation{Fakult\"at f\"ur Mathematik, Universit\"at Bielefeld, 
  Universitätsstr.~25, 33615 Bielefeld, Germany}
\author{Johannes Roth}
\author{Hans-Rainer Trebin}
\affiliation{\itap}

\begin{abstract}
  We extend the Wolf direct, pairwise $r^{-1}$ summation method with
  spherical truncation to dipolar interactions in silica. The
  Tangney-Scandolo interatomic force field for silica takes regard of 
  polarizable oxygen atoms whose dipole moments are determined by 
  iteration to a self-consistent solution. With Wolf summation, the 
  computational effort scales linearly in the system size and can easily 
  be distributed among many processors, thus making large-scale
  simulations of dipoles possible. The details of the implementation
  are explained. The approach is validated by estimations of the
  error term and simulations of microstructural and thermodynamic
  properties of silica.

\bigskip
\noindent
Copyright (2010) American Institute of Physics. This article may be
downloaded for personal use only. Any other use requires prior
permission of the author and the American Institute of Physics. 
\end{abstract}

\maketitle

\section{Introduction}
\label{sec:intro}

Silica is by far the most abundant mineral in the earth's
crust\cite{itapdb:Iler1979}. This makes it an interesting system to
study in simulation. Additionally, SiO$_2$ shows a wide range of
crystalline structures depending on temperature and pressure, and it
can also be solidified as a glass. Although there have been enormous
advances in \emph{ab initio} simulations of silica\cite{itapdb:Karki2007}, 
many effects are inaccessible due to length and time scale
restrictions of these models.  
For large-scale atomistic simulations, a high-quality model 
of the interactions, a so-called effective potential or force
field, is essential.

Many attempts to parameterize the interactions in silica have been made
in the past thirty years, with various levels of computational intensity
and accuracy. Some of the earlier potentials are still widely used,
like for example the potential of van Beest, Kramer, and van Santen (BKS)
\cite{itapdb:Beest1990}, a pure pair
potential with fixed charges and short-range corrections. However, it
is believed that many-body effects are important for correctly describing 
bond angles and bond-bending vibration frequencies in network-forming
glasses like SiO$_2$.\cite{itapdb:Wilson1996,itapdb:Wilson2000} The potential
model of Tangney and Scandolo\cite{itapdb:Tangney2002} (TS) treats
the oxygen atoms as polarizable. The dipole moments of these atoms are
determined self-consistently from the local electric field, with
short-range corrections to the
polarization.%
\cite{itapdb:Rowley1998}
A more detailed description of the TS potential is given in Sec.~\ref{sec:ts}.

A comparison of various silica force fields showed\cite{itapdb:Herzbach2005} 
that the polarizable ion model of TS yields
significantly better results for many properties compared to the BKS
potential, while still leaving room for improvement. 
In a recent study by Paramore \etal\cite{itapdb:Paramore2008}, 
attempts to map the implicit many-body
effects in the TS model to pure pairwise interactions did not lead to
an accurate potential. This confirms that polarization effects are indeed
necessary for a proper description of SiO$_2$.

In all potential models discussed above, the ions carry some charge
$q_{i}$ and interact with a Coulomb potential. This leads to the
classical Madelung problem:\cite{itapdb:Madelung1918} determining the
energy of a condensed system with a pairwise $r^{-1}$ interaction. The
convergence properties of the resulting sum require a special
treatment, and a number of methods to evaluate the pairwise $r^{-1}$
sum have evolved, with the Ewald method\cite{itapdb:Ewald1921} as the
best-known. There, rapid convergence for the total Coulomb
energy of a set of $N$ ions with charge $q_{i}$ at positions
$\ve{r}_{i}$ that are part of an infinite system of point charges,
\begin{equation}
  \label{eq:Coulomb}
  E^{\text{tot}} = \frac{1}{2} \sum\limits_{i=1}^{N}
  \sum\limits_{j\neq i=1}^{\infty}\frac{q_{i}q_{j}}{r_{ij}},
\end{equation}
(where $\ve{r}_{ij}=\ve{r}_{j}-\ve{r}_{i}$ and
$r_{ij}=\left|\ve{r}_{ij}\right|$) is assured by a mathematical
trick. Firstly, structural periodicity of linear size $L$ is
artificially imposed on the 
system, and in the resulting expression a decomposition of unity of 
the form
\begin{equation}
  \label{eq:unity}
1=\erfc\!\left(\kappa r\right)+\erf\!\left(\kappa r\right)  
\end{equation}
is inserted. The error function is defined as
\begin{equation}
\erf\!\left(\kappa
  r\right):=\frac{2}{\sqrt{\pi}}\int\limits_0^{\kappa
  r}\!dt\,e^{-t^2}.
\end{equation}
The Ewald splitting parameter $\kappa$ controls the 
distribution of energy contributions between the two terms. Thus,
Eq.~\eqref{eq:Coulomb} can be written as
\begin{equation}
\label{eq:Ewald}
\begin{split}
E^{\text{tot}} = & \frac{1}{2} \sum\limits_{i=1}^{N}
\sum\limits_{j=1}^{N} \sum\limits_{\ve{n}=\ve{0}}^{\infty}\,'
\frac{q_{i}q_{j}}{|\ve{r}_{ij}+\ve{n}L|} \left[\erfc(\kappa
  |\ve{r}_{ij}+\ve{n}L|)\right.  \\ 
&\left.+\erf(\kappa|\ve{r}_{ij}+\ve{n}L|)\right] ,
\end{split}
\end{equation}
where the sum over periodic images $\ve{n}$ is primed to indicate that
the $i=j$ term is to be omitted for $\ve{n}=\ve{0}$. Taking the
Fourier transform of the error-function  
expression only, but not of
the complementary error-function term, one can convert the
conditionally convergent total energy Eq.~\eqref{eq:Coulomb} into the
sum of real-space and reciprocal-space contributions
$E_{\ve r}^{\text{tot}}$ and $E_{\ve k}^{\text{tot}}$ where each of these converges
rapidly. The downside to the Ewald summation method is the scaling of
the computational effort with the number of particles in the
simulation box: Even when the balance between real- and
reciprocal-space contributions controlled by $\kappa$ is optimized, the
computational load increases at best as
$O(N^{3/2})$.\cite{itapdb:Fincham1994} For large-scale simulations
with millions of atoms, this is insufficient. Additionally, the
Ewald technique is limited to periodic systems. In recent years,
alternative simulation techniques that show better scaling properties
have been developed, among them mesh-based methods or fast multipole
methods.\cite{itapdb:Gibbon2002} %
The linear scaling, however, comes with considerable overhead. In
contrast, Wolf \etal\cite{itapdb:Wolf1999} proposed a direct summation
technique with linear scaling ($O(N)$) for Coulomb interactions,
that can easily be implemented in standard Molecular Dynamics (MD)
codes. This so called Wolf summation takes into account the physical 
properties of the systems under study.

To this end, one looks at the Fourier transform of the error function
term of Eq.~\eqref{eq:Ewald} 
\begin{equation}
  \label{eq:Ektot}
E_{\ve k}^{\text{tot}} = 
\frac{2\pi}{L^{3}} \sum_{\ve k\ne \ve 0}\sum_{i,j}
       q_{i}q_{j}e^{i\ve k\cdot(\ve r_{j}-\ve r_{i})}
       \frac{\exp(-\frac{|\ve k|^{2}}{4\kappa^{2}})}{|\ve k|^{2}}
     - \frac{\kappa}{\pi^{1/2}}\sum_{i}q_{i}^{2},
\end{equation}
where the self term ($\ve{n} = \ve{0}$ and $i = j$) is now included in
the summation and subtracted again separately. Eq.~\eqref{eq:Ektot}
can be rewritten as
\begin{align}
  \label{eq:Ek_S}
 E_{\ve k}^{\text{tot}} &= \sum_{\substack{\ve k\ne\ve 0\\ |\ve k|<k_{c}}}
      S(k)\frac{\exp(-\frac{|\ve k|^{2}}{4\kappa^{2}})}{|\ve k|^{2}}
     - \frac{\kappa}{\pi^{1/2}}\sum_{i}q_{i}^{2} ,\\
\intertext{where $S(k)$, with $k=|\ve{k}|$, is the charge structure factor}
 \label{eq:csf}
 S(k) &= \frac{2\pi}{L^{3}} \left| \sum_{j} q_{j}
       \exp(i\ve{k}\cdot \ve{r}_{j})\right|^{2}.
\end{align}
The charge structure factor is the Fourier transform of the
charge-charge autocorrelation function.

In the systems of interest here, there are no long-range charge
fluctuations; the charges form a cold dense plasma, screening each other.
This means that for small wave vectors $k$, the charge structure
factor is also small. If one now chooses a sufficiently small Ewald
parameter $\kappa$, the reciprocal-space contribution can be neglected
altogether. As $\kappa$ is linked to the real-space cut-off $r_c$, 
however, this might require a cut-off radius which is substantially 
larger than the range of traditional short-range interactions like in 
metals.

Concurrently, Wolf \etal\ also motivated a continuous and smooth
cut-off of the remaining screened Coulomb potential
$V(r_{ij})=q_iq_j\erfc(\kappa r_{ij}) r_{ij}^{-1}$ at a cut-off radius
$r_c$. The authors stated that shifting the pair potential so that it goes
to zero smoothly at $r=r_c$ is equivalent to neutralizing the surface
charge in a spherically truncated system. The strong fluctuations in
the surface charge with varying $r_c$ inhibit the convergence to the 
true Madelung energy with increasing $r_c$. The combination of 
(i) shifting the potential so that it vanishes smoothly at the cut-off, and 
(ii) damping the Coulomb potential to reduce the required cut-off radius,
but only so weakly that the reciprocal-space term can still be neglected,
is called Wolf summation.

To evaluate the TS potential with the Wolf direct summation technique,
one first has to extend the formalism to the treatment of dipolar
interactions. How this is done is shown in Sec.~\ref{sec:wolfdip}. We
provide an estimate of the errors made by the approximation in
Sec.~\ref{sec:conv}. The directly summed TS potential was
implemented in the (limited range) MD code IMD,\cite{itapdb:Stadler1997a} 
and various observables were determined and
compared to the original TS implementation with full Ewald summation
(Sec.~\ref{sec:res}). Finally, we sum up the results in
Sec~\ref{sec:diss}, where also an outlook is given.

\section{Wolf summation of dipole contributions}
\label{sec:wolfdip}

\subsection{Tangney-Scandolo potential model}
\label{sec:ts}

In the TS\cite{itapdb:Tangney2002} force field,
there are two contributions to the potential energy of a system: a
pairwise potential of Morse-Stretch form, and the electrostatic
interactions between charges and induced dipoles on the oxygen
atoms. The dipole moments depend on the local electric
field at the respective atomic sites, which in turn is determined by
the arrangement of charges and dipoles. This implies that a
self-consistent solution must be found.

Tangney and Scandolo propose an iterative solution for the dipole
moments, so that the dipole moment $\ve{p}_i^n$ on atom $i$ in iteration
step $n$ is
\begin{equation}
  \label{eq:scloop}
  \ve{p}_i^n=\alpha\ve{E}(\ve{r}_i;\{\ve{p}_j^{n-1}\}_{j=1,N},
  \{\ve{r}_j\}_{j=1,N}) +  \ve{p}_i^{\text{SR}},
\end{equation}
where $\alpha$ is the polarizability of atom $i$ and $\ve{E}(\ve{r}_i)$
the electric field at position $\ve{r}_i$, which is calculated from
the dipole moments (and charges) in the previous iteration step. 
The short-range 
dipole moment $\ve{p}_i^{\text{SR}}$ is the contribution induced by
short-range repulsive forces between anions and cations, that TS
included following Rowley \etal\cite{itapdb:Rowley1998}
Starting from initial electric field strengths $\ve{E}^0(\ve{r})$
extrapolated from the previous three time steps, Eq.~\eqref{eq:scloop}
is iterated until convergence is achieved for each MD time step. 

The parameters of the TS potential were determined solely from
\emph{ab initio} results with the Force Matching
method\cite{itapdb:Ercolessi1994}. There, the potential is
parameterized using first principles values of forces, stresses and
energies in series of reference structures.

\subsection{Smooth cut-off}
\label{sec:smooth}

For MD with limited-range interactions, the potentials 
and their first derivatives must go to zero continuously at a cut-off 
radius $r_c$; otherwise, atoms crossing this threshold might get unphysical
kicks. For the Morse-Stretch pair potential, this is generally not
problematic, as it decays with $r_{ij}$ fast enough. In MD, 
following Wolf \etal,\cite{itapdb:Wolf1999} the
potential $U_{\text{MS}}(r_{ij})$ is replaced by 
\begin{equation}
  \label{eq:smooth}
  \tilde{U}_{\text{MS}}(r_{ij}) =
  U_{\text{MS}}(r_{ij}) - U_{\text{MS}}(r_c) -
  (r_{ij}-r_c) U'_{\text{MS}}(r_c),  
\end{equation}
where a prime denotes a derivative with respect to $r$.

The other functions used in the TS model have a general $r$ dependence of
the form $r^{-n}, n\in\{1,2,3\}$. Especially the Coulomb energy with
its $r^{-1}$ dependency cannot simply be cut off without a treatment
as in Eq.~\eqref{eq:smooth}, for otherwise the energy of the system 
would fluctuate strongly with $r_c$, without convergence to the proper
value. But even with a smooth cut-off (\ref{eq:smooth}), with which
the Coulomb energy does converge, a rather large cut-off radius would be 
required to make shifting of the potential negligible. Fortunately, 
the Wolf direct summation method\cite{itapdb:Wolf1999} includes a weak 
exponential damping of the Coulomb potential by
$\erfc\!\left(\kappa r\right)$. Such a damped potential can be cut off
smoothly at a much smaller radius $r_c$ without affecting the result. All
integer powers of $r^{-1}$ are treated in a way to conserve the
differential relationship between the functions, i.e. the damped
functions are
\begin{align}
  \label{eq:rminus1}
  r^{-1} \rightarrow &r^{-1} \erfc(\kappa r)=: f_{-1}(r), \\
  \label{eq:rminus2}
  \begin{split}
  r^{-2} = -\frac{d( r^{-1})}{d r} \rightarrow &-\frac{d(r^{-1}
    \erfc(\kappa r))}{d r}\\
&=r^{-2}\erfc(\kappa r) - \frac{2\kappa\exp(-\kappa^2 r^2)}{\sqrt{\pi}r}\\
&=: f_{-2}(r).
\end{split}
\end{align}
This procedure is also required to conserve the energy during an MD
simulation, as discussed in more detail in Sec.~\ref{sec:energy}.

The damped potentials are then shifted to zero and zero derivative at
the cut-off radius, as in (\ref{eq:smooth}). This allows for 
limited-range MD simulations with a standard MD code. 
The computational effort of such a simulation
scales linearly in the number of particles (as the number of
interactions that need to be evaluated per particle does not increase
with the number of particles), but scales roughly with $O(r_c^3)$.

\subsection{Energy conservation}
\label{sec:energy}
In MD simulations, the energy is conserved, if the
forces on the particles are exactly equal to the gradient of the
potential energy with respect to the atomic coordinates. Otherwise,
the energy might oscillate or even drift off if not controlled by a
thermostat. In standard MD simulations, the requirement is usually
automatically fulfilled: The forces are calculated as the derivative
of the potential, which depends directly on the atomic positions. In the
TS model, there is also an indirect dependence, as the potential is
also a function of the dipole moments:
\begin{equation}
  \label{eq:potdep}
  U=U(\{\ve{r}_i\}, \{\ve{p}_i(\{\ve{r}_j\})\}).
\end{equation}
This would in principle lead to an extra contribution to the
derivative of the potential,
\begin{equation}
\label{eq:potder}
\frac{d U}{d \{\ve{r}_i\}}=\frac{\partial U}{\partial \{\ve{r}_i\}} +
\frac{\partial U}{\partial \{\ve{p}_i\}}
\frac{\partial\{\ve{p}_i\}}{\partial \{\ve{r}_j\}}, 
\end{equation}
which would be practically impossible to be determined effectively. Luckily,
if the dipole moments are iterated until convergence is reached, we are 
at an extremal value of the potential energy, with %
$\partial   U/\partial \{\ve{p}_i\}=0$, and so this part need not be evaluated.
Imperfections in convergence may lead to a drift in the energy, 
however, as was already observed by Tangney and
Scandolo.\cite{itapdb:Tangney2002}

When applying the Wolf formalism to the TS potential, another issue
arises concerning the conservation of energy. It can most
easily be explained with a simple one-dimensional example. Given are two
oppositely charged point charges $\pm q$ at a mutual distance $r$. If the
negatively charged one is polarizable with polarizability $\alpha$, it
will get a dipole moment %
$p=\alpha q / (k r^2)$, with
$k=4\pi\epsilon_0$. This leads to a total interaction energy
\begin{align}
  \label{eq:eng2at}
  U &= \underbrace{-2\cdot\frac{1}{2}\frac{1}{k}\frac{q^{2}}{r}}_{q-q}
  - \underbrace{2\cdot\frac{1}{2}\frac{q}{k}\frac{p}{r^{2}}}_{q-p}
  + \underbrace{\frac{1}{2}\frac{p^{2}}{\alpha}}_{\text{dipole}},\\
  \intertext{from which it follows that} \frac{\partial U}{\partial
    p}&=-\frac{1}{k}\frac{q}{r^{2}}+
  \underbrace{\frac{p}{\alpha}}_{=\frac{1}{k}\frac{q}{r^{2}}}=0.
\end{align}
Here, $q-q$ denotes the Coulomb interaction between charges, $q-p$ the
interactions between charge and dipole, and the last term is the
dipole energy. When we now damp and cut off the interactions, we
replace the $r^{-1}, r^{-2}$ functions by their damped and smoothed
counterparts $\tilde{f}_{-1}(r), \tilde{f}_{-2}(r)$. If energy
conservation is to be maintained, the differential relation between the
$\tilde{f}_{-n}$ must be the same as for the $r^{-n}$:
\begin{equation}
  \label{eq:dampf}
  \frac{d \tilde{f}_{-1}(r)}{d r}=-\tilde{f}_{-2}(r).
\end{equation}
As a consequence, the first two derivatives of the smoothed damped
Coulomb potential must be zero at $r_c$. 

In MD simulation it is computationally advantageous to represent pair
potential functions internally as functions of $r^2$, and their
derivative as %
$f\hat{'}:=r^{-1}df/dr$. The damped
Coulomb potentials $f_{-1}$ and $r^{-1} f_{-2}(r)$ in their
smoothly cut off version become 
\begin{align}
  \label{eq:smdc}
  \begin{split}
  \tilde{f}_{-1}(r^2) &=f_{-1}(r^2)-f_{-1}(r_c^2) \\ &\quad-
  \left.\tfrac{1}{2}f\hat{'}_{-1}(r^2)\right|_{r^2=r_c^2} (r^2-r_c^2)  \\ & \quad-
  \left.\tfrac{1}{8}f\hat{'}\hat{'} _{-1}(r^2) \right|_{r^2=r_c^2}(r^2-r_c^2)^{2}
  \end{split}
  \intertext{and}  
  \label{eq:smdd}
  \begin{split}
  \quad\tfrac{1}{r} \tilde{f}_{-2}(r) &= 
  \tfrac{1}{r}f_{-2}(r^2)
  -\left.f\hat{'} _{-1}(r^2)
  \right|_{r^2=r_c^2}  \\
 &\quad-\left.\tfrac{1}{2}f\hat{'}\hat{'} _{-1} (r^2)\right|_{r^2=r_c^2} (r^2-r_c^2).
  \end{split}
 \end{align}
In this way, Wolf summation can be applied to dipolar interactions in
the TS potential model. In Sec.~\ref{sec:conv} we will discuss
why this approximation is physically justified.

\subsection{Implementation}
\label{sec:impl}
The ITAP Molecular Dynamics (IMD) package\cite{itapdb:Stadler1997a} 
is a flexible, highly scalable MD code for limited-range interactions,
providing linear scaling up to thousands of CPUs. For finite-range
interactions, the number of potential interaction partners of an
atom is uniformly bounded. In order to reach linear scaling in the
number of atoms, it is essential to find these interaction partners
efficiently. IMD uses a combination of link-cells and neighbor lists, 
where the former are used to compute the latter in an efficient way.
Since Wolf summation requires a relatively large cut-off radius, 
these neighbor list can get fairly big, but on today's machines this
is not a problem. Parallelization is done via a fixed geometric domain 
decomposition, where each CPU gets an equal block of material. For
the force computation, atoms at the surface of a block are exchanged 
with the neighboring CPUs.

All potential functions used in IMD are tabulated, even if some of
these functions may be specified by giving the parameters of an
analytic formula. In that case, potential tables are constructed
from the analytic formula in a pre-processing step. During the 
simulation loop, the functions are then evaluated by table lookup 
and interpolation. This has proven to be the most flexible and
efficient scheme, allowing also for very complicated potential 
functions. For all potential functions depending on the radius,
care is taken that they vanish smoothly at the cut-off radius,
along with their first derivative.

In contrast to other interactions implemented in IMD, the TS potential
requires a self-consistency loop within each time step, during which
the dipole strengths of the oxygen atoms are determined. Before entering
this loop, the ``static'' contributions $\ve{E}_{\text{stat}}$ to the on-site
electric field caused by the charges of anions and cations, and the
short-range dipole contributions $\ve{p}_i^{\text{SR}}$ are calculated
and stored.  For the ``induced'' part of the electric field 
$\ve{E}_{\text{ind}}$, which is generated by the oxygen dipoles, 
Eq.~\eqref{eq:scloop} is then iterated until convergence is achieved. 
The iteration starts from an extrapolation of the local
electric field at the previous three MD time steps. To improve the
convergence of Eq.~\eqref{eq:scloop}, $\ve{E}_{\text{ind}}^n$ is
modified after each iteration step $n$ to include a small part $c$ from the
previous iteration,
\begin{equation}
  \label{eq:dampE}
  \ve{E}_{\text{ind}}^n \rightarrow (1-c) \ve{E}_{\text{ind}}^n +c\, 
    \ve{E}_{\text{ind}}^{n-1}.
\end{equation}
This damps the self-consistency loop and thus suppresses overshooting
the optimal solution and subsequent oscillations. For optimal
performance, a value of $c=0.2$ was used.

Convergence is achieved, when the root mean square deviation of all
Cartesian dipole moment components between two iterations is less
than a user-specified tolerance (given in units of the dipole
moment). While a larger tolerance will reduce the iteration steps to
convergence, it 
will also introduce a larger error in the energy conservation, which might
lead to a temperature drift in microcanonical simulations. In
practice, a convergence limit smaller than $10^{-6}$ \AA$e$\
(with elementary charge $e$) will not
lead to further improvement. With this tolerance, about five iterations
steps are typically needed per MD step.

In a parallel simulation, each CPU deals with a block of material.
For the parallel evaluation of the energies and forces, at each MD step
the types and positions of atoms near the surface of a block are first 
communicated to the neighboring CPUs. Each CPU can then perform a
part of the energy and forces computation locally. As each force 
is computed only once, certain force and energy contributions have then
to be communicated back to the home CPU of the corresponding atom, where
it is added up. This scheme is valid for all finite range interactions.
Since only communication between neighboring CPUs is necessary, the
scheme is highly scalable. 

For the TS potential the procedure is very similar, except that now
there are additional data to be communicated. In each step of the
self-consistency loop for the induced dipoles, the electric fields and
dipole moments of atoms at the surface must be distributed to the 
neighboring CPUs, and collected again after they have been updated.
There are several additional communication steps for each MD step,
but these are of the same kind as for other short-range interactions
(to neighbor CPUs only), and the balance between communication and
computation is not affected. For this reason, simulations with the
TS potential will scale as well as with other short-range potentials.

\section{Convergence and Error Estimation}
\label{sec:conv}

\subsection{Formal Analysis}

The total interaction energy of $N$ dipole moments $\ve{p}_i$ at
positions $\ve{r}_i$ is given by the expression 
\begin{equation}
E^{\text{tot}}=
-\frac{1}{2}\sum\limits_{\genfrac{}{}{0pt}{}{i,j}{i\neq j}}^{N}
\,\ve{p}^t_i\left(\ve{\nabla}\otimes\ve{\nabla}\right)
\left(\frac{1}{r_{ij}}\right)\ve{p}_j, 
\end{equation}
with $\ve{r}_{ij}:=\ve{r}_i-\ve{r}_j$ and
$r_{ij}:=|\ve{r}_{ij}|$. Imposing structural periodicity and inserting
a decomposition of unity of the form 
\begin{equation*}
\tag{\ref{eq:unity}}
1=\erfc\!\left(\kappa r\right)+\erf\!\left(\kappa r\right),
\end{equation*}
where $\kappa$ is again the Ewald splitting parameter, 
we can rewrite above equation as
\begin{multline}
\label{etot01}
E^{\text{tot}}=
-\frac{1}{2}\sum\limits_{i,j}^{N}\sum\limits_{\ve{n}=\ve{0}}^{\infty}
\text{$'$}\,\ve{p}^t_i\left(\ve{\nabla}\otimes\ve{\nabla}\right)\\
\cdot\left(\frac{\erfc\!\left(\kappa|\ve{r}_{ij}+\ve{n}L|\right)
+\erf\!\left(\kappa|\ve{r}_{ij}+\ve{n}L|\right)}{|\ve{r}_{ij}+\ve{n}L|} 
\right)\ve{p}_j. 
\end{multline}
The total energy splits into a real- and a reciprocal-space
part: 
\begin{equation}
E^{\text{tot}}=E^{\text{tot}}_{\ve{r}}+E^{\text{tot}}_{\ve{k}}
\end{equation}
Since we later intend to neglect the reciprocal-space term for the 
Wolf summation, we
are interested in the contribution of $E^{\text{tot}}_{\ve{k}}$. For
its $\ve{k}$-behavior we have to take the Fourier transform of 
\begin{equation}
E^{\text{tot}}_{\ve{k}}=
-\frac{1}{2}\sum\limits_{i,j}^{N}\sum\limits_{\ve{n}=\ve{0}}^{\infty} 
\,\ve{p}^t_i\left(\ve{\nabla}\otimes\ve{\nabla}\right)
\left(\frac{\erf\!\left(\kappa|\ve{r}_{ij}+\ve{n}L|\right)}%
{|\ve{r}_{ij}+\ve{n}L|}\right)\ve{p}_j.  
\end{equation}
The prime has been omitted, since the self term (for $\ve{n}=\ve{0}$ and
$i=j$) is now finite. Because of the three-dimensional periodicity the above 
expression can be expanded into a Fourier series: 
\begin{equation}
\label{etotrezfou}
\widetilde{E}^{\text{tot}}_{\ve{k}}=\frac{2\pi
  Ne^2}{V}\sum\limits_{\ve{k}\neq\ve{0}}^{\infty}\,
\ve{k}^t\te{Q}(\ve{k})\ve{k}\;
\dfrac{\exp\!\left(-k^2/4\kappa^2\right)}{k^2},  
\end{equation}
where $V$ is the volume of the simulation cell and $\te{Q}(\ve{k})$ 
the dipole structure factor
\begin{equation}
\te{Q}(\ve{k}):=\frac{1}{Ne^2}\sum\limits_{i,j}^{N}\,
 \ve{p}_i\otimes\ve{p}_j\,e^{i\ve{k}\cdot\ve{r}_{ij}},
\end{equation}
with the normalization factor $1/\sqrt{Ne^2}$, where $e$ denotes 
the elementary charge. As we can see in
Eq.~\eqref{etotrezfou}, the large $\ve{k}$ contributions to
$\widetilde{E}^{\text{tot}}_{\ve{k}}$ tend to zero rapidly,
whereas the small $\ve{k}$ contributions are governed by the
behavior of $\te{Q}(\ve{k})$, which is expected to vanish 
as $k \rightarrow 0$.

\subsection{Discussion}

To legitimate the neglecting of the reciprocal-space term for the Wolf
summation we have simulated liquid silica with 4896 atoms, where we get no
spontaneous polarization as a first result. The total dipole moment
is $p = 6.76\cdot10^{-29}$\, Cm, which is insignificantly small compared
to a fully polarized system and thus can be taken as a
fluctuation. All values which are calculated in the course of the
simulation are time-averaged over the full simulation time of
one picosecond.  

\begin{figure}
\centering
\includegraphics{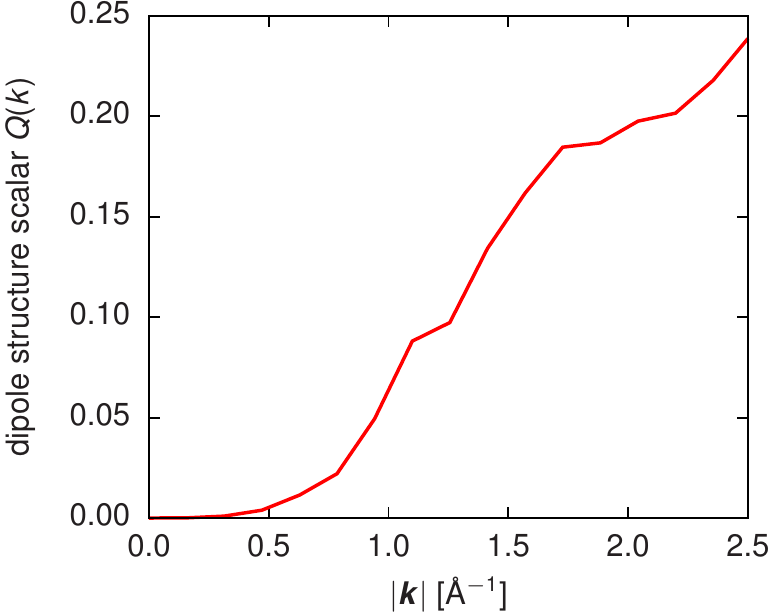}
\caption{$k$-dependence of the dipole structure scalar $Q(k)$. For small $k$,
  the dipole structure factor is negligible.}
\label{fig:dipolestruc}
\end{figure}

To analyze the $k \rightarrow 0$ behavior we calculated the 
dipole structure scalar, 
\begin{equation}
Q(k) = \langle \ve{k}^t\te{Q}(\ve{k})\ve{k}\ \rangle_S,
\end{equation}
where the angular brackets indicate an average over a spherical shell
$S$ with width $\Delta k$ centered at constant $|\ve{k}|=k$.  Note
that for a for a periodic system $\te{Q}$ is not a continuous
function, but a discrete set, consisting of all reciprocal space
vectors. Hence the average over the spherical shell is necessary. Fig.\
\ref{fig:dipolestruc} shows the dipole structure scalar in liquid
silica simulations. For small absolute values of $k$, $Q(k)$ goes to
zero.

\begin{figure}
\centering
\includegraphics{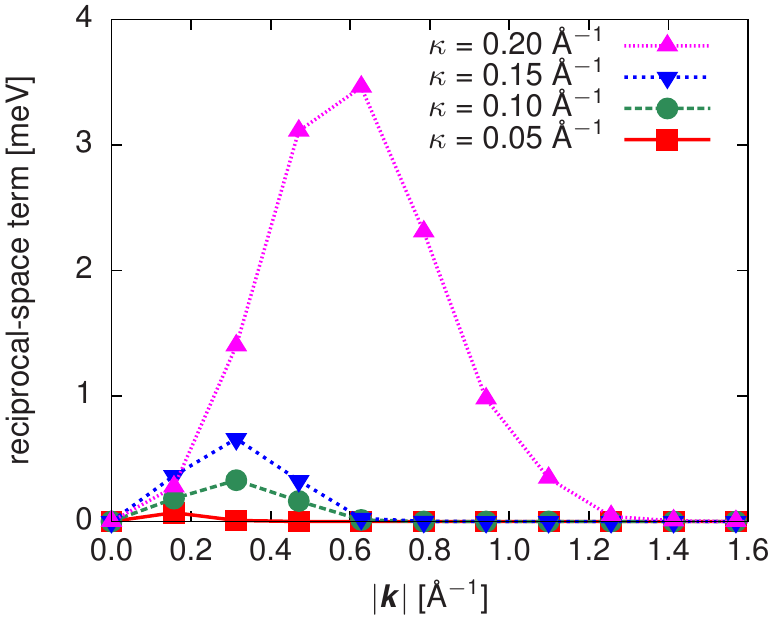}
\caption{$k$-dependence of the reciprocal-space term 
  $\widetilde{E}_{\ve{k}}(k)$ for different Ewald 
  splitting parameters $\kappa$. The $k \rightarrow 0$ 
  behavior of $\widetilde{E}_{\ve{k}}(k)$ is governed by $Q(k)$, 
  which results in negligible contributions of the small $k$-values
  to the total energy.} 
\label{fig:recienergy}
\end{figure}

Fig.\ \ref{fig:recienergy}
shows the $k$-dependence of the reciprocal-space term, 
\begin{equation}
\widetilde{E}_{\ve{k}}(k) = \frac{2\pi Ne^2}{V}\, Q(k)\,
\dfrac{\exp\!\left(-k^2/4\kappa^2\right)}{k^2}, 
\label{kdepetot}
\end{equation}
for different Ewald splitting parameters $\kappa$ (again averaged
over a spherical shell). As
mentioned above, due to the exponential damping, large-$k$ 
contributions are negligibly small, whereas the small-$k$ values 
are governed by the behavior of $Q(k)$ as $k \rightarrow 0$.  

\begin{figure}
\centering
\includegraphics{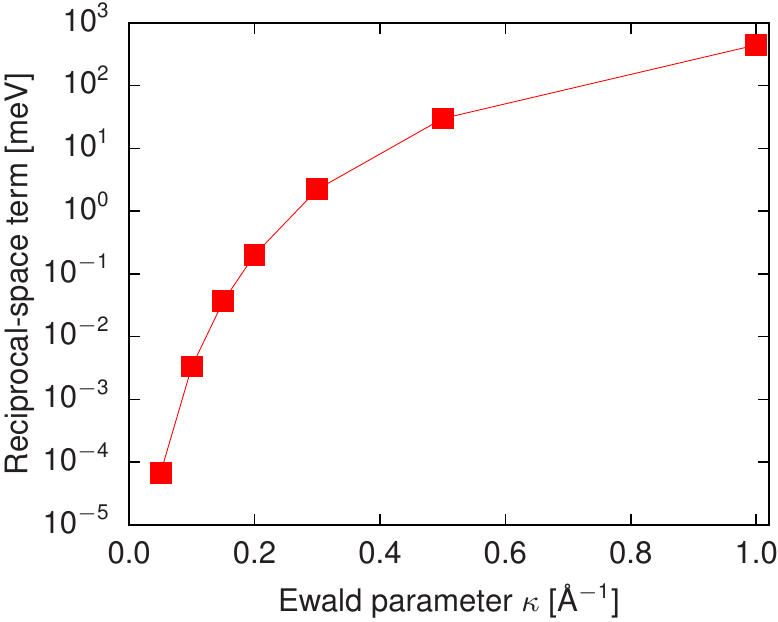} 
\caption{Logarithmic plot of the reciprocal-space term
  $\widetilde{E}^{\text{tot}}_{\ve{k}}$ for different Ewald splitting
  parameters $\kappa$. For sufficiently small $\kappa$, there is no
  noticeable contribution to the total energy compared to the
  real-space part.}
\label{fig:energylog}
\end{figure}

Finally the sum in Eq.~\eqref{etotrezfou} is evaluated for the given
$k$-mesh with truncation sphere in the reciprocal-space. 
The difference between this approach of a
spherical truncation and the full summation is very small because of
the exponential damping in Eq.~\eqref{kdepetot}, as seen in the
rapid decay of $\widetilde{E}_{\ve{k}}(k)$ for increasing $k$ in
Fig.~\ref{fig:recienergy}. In Fig.~\ref{fig:energylog} the $\kappa$-dependence
of the reciprocal-space term $\widetilde{E}^{\text{tot}}_{\ve{k}}$ is
illustrated in a logarithmic plot. For the chosen damping of $\kappa
= 0.1 \, \text{\AA}^{-1}$ we get 
\begin{equation}
\label{eq:trunc}
\frac{1}{N}\widetilde{E}^{\text{tot}}_{\ve{k}}  = 3.3 \, \mu{\text eV}, 
\end{equation}
which is small compared to the real-space part and can thus
be neglected. 

\section{Results}
\label{sec:res}

The damped and smoothly cut off TS potential was used to
study the same thermodynamic and structural properties the original
authors\cite{itapdb:Tangney2002} examined for the Ewald-summed
potential. 

\subsection{Equation of State and Bonding Properties}

\begin{figure}
  \centering
  \includegraphics{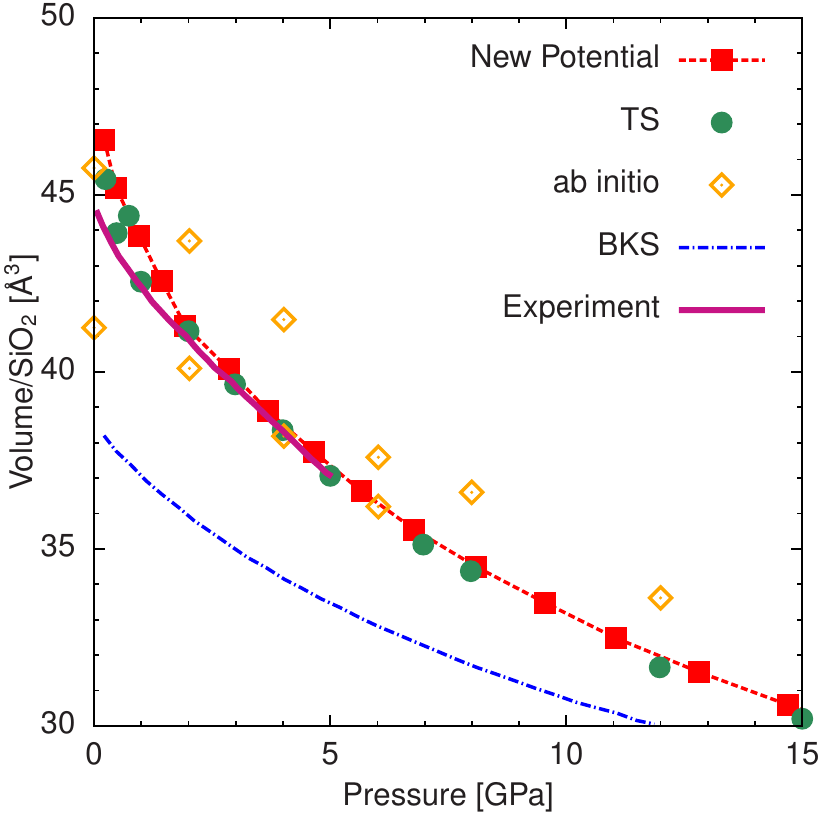}
  \caption{Equation of state of liquid silica for damped and smoothly
    cut off TS potential compared to
    experiment\cite{itapdb:Gaetani1998}, \emph{ab initio} simulations
    and classical simulations with BKS and the full TS
    potential\cite{itapdb:Tangney2002}.}
  \label{fig:eqstate}
\end{figure}

We compare the equation of state of liquid silica at
3100~K to experiments\cite{itapdb:Gaetani1998}, \emph{ab initio}
results and, of course, the full TS potential in
Fig.~\ref{fig:eqstate}. Pressures were obtained 
as averages along constant-volume MD runs of
approximately 10 ps following 10 ps of equilibration and with
simulation cells containing 4896 atoms. We reproduced the good
agreement of the full TS potential with the experimental results; both 
the full TS potential and our damped and smoothly cut off TS potential 
match even better with experiment than the \emph{ab initio}
results. As already mentioned by the original authors\cite{itapdb:Tangney2002} 
the BKS model systematically underestimates the volume by $\approx$ 13\%.
The large scatter of the \emph{ab initio} results can be explained
with the system size and time constraints of this method: especially
for low pressures, the system cannot be equilibrated completely. 

\begin{figure}
  \centering
  \includegraphics{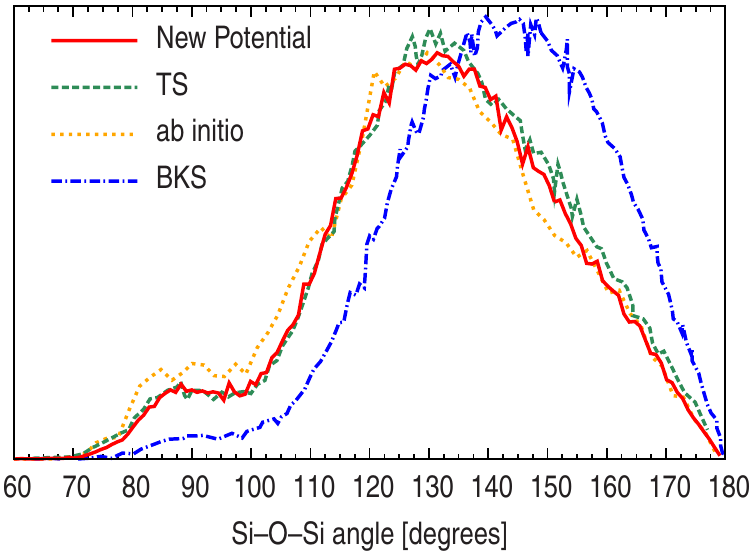}
  \caption{Oxygen centered angle distribution in liquid silica for the 
	new potential compared to simulations with BKS and the full TS
        potential as well as \emph{ab initio}
        calculations.\cite{itapdb:Tangney2002}} 
  \label{fig:angle}
\end{figure}

On a microscopic level, the Si--O--Si angle distribution was determined
from multiple MD simulation runs at 3100 K and various pressures. The
results are shown in Fig.~\ref{fig:angle}, and are in agreement with 
the full TS potential and \emph{ab initio} results. 

\begin{figure}
  \centering
  \includegraphics{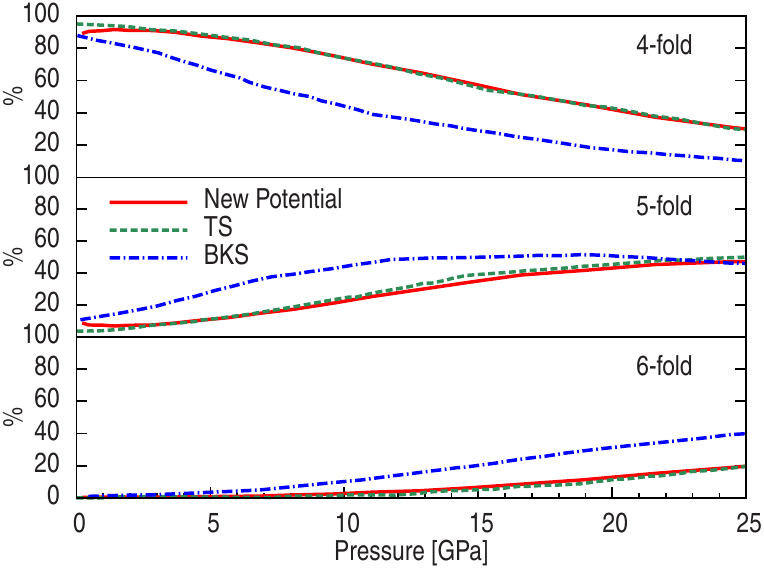}
  \caption{Percentage of $N$-fold coordinated silicon atoms in liquid silica 
at 3100 K as a function of pressure compared to simulations with BKS and 
the full TS potential\cite{itapdb:Tangney2002}.}
  \label{fig:coord}
\end{figure}

In Fig.~\ref{fig:coord} the percentage of $N$-fold coordinated silicon
atoms in liquid silica at 3100 K as a function of pressure is
illustrated. Our results are compared to simulations with the full TS
potential,\cite{itapdb:Tangney2002} which agree rather well with
\emph{ab initio}\cite{itapdb:Barrat1997} results.

To sum up, the equation of state and the bonding properties of liquid
silica, which the original authors\cite{itapdb:Tangney2002} examined
for the Ewald-summed potential, can be reproduced very well by using
the damped and smoothly cut off TS potential, while using dramatically
less CPU time. Due to the linear scaling of computational effort in
the system size, this advantage becomes even more pronounced the
larger the system is.

\subsection{Crystal Structure Data}

We also probed the damped and smoothly cut off TS potential by simulating 
the most important low pressure crystal structures quartz, cristobalite and coesite. The 
relevant equilibrium variables density, Si--O--Si angle and the lattice
parameters at 300 K are given in 
Tables~\ref{tab:quartz}, ~\ref{tab:cristobalite} and
\ref{tab:coesite}. The average relative 
deviation of the data from the experimental results is $\approx$
0.9\%, which is a 
comparatively good agreement. By contrast, the BKS potential differs by
$\approx$ 2.1\% on average. 
Note that simulations with the full TS potential yield a relative deviation of the
parameters that averages at merely $\approx$ 0.7\%. 
This decrease in precision might be countered by redetermining the
parameters for the smoothed and damped TS force field, as we suggest
in Sec.~\ref{sec:diss}.
Additionally, the ordered crystals might be more susceptible
to spontaneous polarization  compared to the liquid, however we could
not confirm this in our simulations.

It should be noted, however, that the TS potential was optimized to
reproduce atomistic properties of liquid SiO$_2$ at 3000 K. For this
reason, its application to low-temperature crystalline systems should
be closely monitored. In the case of cristobalite we found that both
the full TS potential and the smoothly truncated potential
energetically favor a slightly different orientational arrangement of
the fundamental SiO$_4$ tetrahedra at low temperatures, with only
little consequence on quantities given in Tab.~\ref{tab:cristobalite}.

\newcommand{\oneline}[5]{
 #1 & #2 & #3 & #4 & #5  \\}

\begin{table}
\caption{Quartz}
   \begin{tabular}{l d d d d}
\hline\hline
& \multicolumn{1}{c}{Expt.\footnote{Reference~  \onlinecite{itapdb:Levien1980}.}} 
&  \multicolumn{1}{c}{New Potential}
& \multicolumn{1}{c}{ TS\footnote{Reference~ \onlinecite{itapdb:Tangney2002}.\label{fn:t1}} }
& \multicolumn{1}{c}{ BKS\footref{fn:t1}} \\
\hline
\oneline{$a$ (\AA)}            {4.916}{4.872}{4.925}{4.941}
\oneline{$c$ (\AA)}            {5.405}{5.359}{5.386}{5.449}
\oneline{$\rho$ (g/cm$^{3}$)}     {2.646}{2.718}{2.665}{2.598}
\oneline{Si--O--Si ($^{\circ}$)}                         {143.7}{142.1}{144.5}{148.1}
\hline\hline
\end{tabular}
\label{tab:quartz}
\end{table}

\begin{table}
\caption{Cristobalite}
   \begin{tabular}{l d d d d}
\hline\hline
& \multicolumn{1}{c}{Expt.\footnote{Reference~  \onlinecite{itapdb:Schmahl1992}. }}
& \multicolumn{1}{c}{New Potential }
& \multicolumn{1}{c}{ TS\footnote{Reference~ \onlinecite{itapdb:Tangney2002}.\label{fn:t2}} }
& \multicolumn{1}{c}{ BKS\footref{fn:t2}} \\
\hline
\oneline{$a$ (\AA)}            {4.969}{5.015}{4.936}{4.920}
\oneline{$c$ (\AA)}            {6.925}{6.999}{6.847}{6.602}
\oneline{$\rho$ (g/cm$^{3}$)}     {2.334}{2.268}{2.412}{2.515}
\oneline{Si--O--Si ($^\circ$)}                         {146.4}{147.1}{144.0}{143.9}
\hline\hline
\end{tabular}
\label{tab:cristobalite}
\end{table}

\begin{table}
\caption{Coesite}
   \begin{tabular}{l d d d d}
\hline\hline
& \multicolumn{1}{c}{Expt.\footnote{Reference~  \onlinecite{itapdb:Levien1981}.} }
& \multicolumn{1}{c}{New Potential} 
& \multicolumn{1}{c}{ TS\footnote{Reference~ \onlinecite{itapdb:Tangney2002}.\label{fn:t3}} }
& \multicolumn{1}{c}{ BKS\footref{fn:t3}} \\
\hline
\oneline{$a$ (\AA)}            {7.136}{7.123}{~7.165~}{~7.138~}
\oneline{$b$ (\AA)}            {7.174}{7.161}{7.162}{7.271}
\oneline{$c$ (\AA)}            {12.369}{12.347}{12.377}{12.493}
\oneline{$\beta$ ($^\circ$)}                        {120.34}{120.34}{120.31}{120.76}
\oneline{$\rho$ (g/cm$^{3}$)}     {2.921}{2.940}{2.933}{2.864}
\oneline{Si--O--Si ($^\circ$)}                         {143.6}{144.2}{144.0}{150.5}
\hline\hline
\end{tabular}
\label{tab:coesite}
\end{table}

\section{Conclusion}
\label{sec:diss}

In this work, we have demonstrated that the advantages of the
TS polarizable force field can be captured and
reproduced in MD simulations with a strictly finite
interaction range. To this end, we have shown that the Wolf summation
technique, i.e.\ smoothly cutting off the damped long range
real space part of the electrostatic interaction, and neglecting the 
reciprocal space part altogether, is justified for the TS
dipolar force field for silica. With a suitably large real space
cut-off, the errors in the forces and energies are acceptable for the
systems of interest. This can also be seen in simulation results: Our
Wolf-summed TS potential can reproduce the experimental and \emph{ab
  initio} structural properties of silica reasonably well compared to
the full TS interaction.

By omitting the reciprocal space contribution, simulations with our
potential can be performed with a standard finite-range MD code 
like IMD. Thus, it can profit from the linear scaling of
computational effort with system size common to this
method. Similarly, the calculations can easily and efficiently be
parallelized, opening the door to large-scale calculations impossible
with the standard Ewald summation technique. Moreover, once the 
reciprocal space part can be neglected, there is no longer any need
for periodic boundary conditions. It has been shown that Wolf summation
performs very well also for open or mixed boundary conditions,%
\cite{itapdb:Wolf1999} opening up a wealth of new possibilities.

As a rule of thumb, the real space cut-off radius required for 
Wolf summation has been estimated as about five times the largest
nearest neighbor distance of opposite charges in the system.%
\cite{itapdb:Demontis2001} For silica, this amounts to a moderate
value of about 8\,\AA. But even with a more conservative choice of 
10\,\AA, for more accurate simulations, for a system with 4896 atoms 
we obtained a speedup of more than two orders of magnitude compared 
to the original code of Tangney and Scandolo. Also this performance
increase makes the new method very interesting, and opens up new
possibilities.

The original TS potential parameters were optimized for the full Ewald
treatment of long-range interactions. Redetermining the parameters for
the smoothed and damped TS force field with the actual cutoff used in
simulation might improve the potential further. Additionally, using a
more flexible short-range interaction than the Morse-Stretch potential
suggested by Tangney and Scandolo could lead to even better results.
We plan to implement the TS polarizable oxide potential
model in our Force Matching code \emph{potfit}\cite{itapdb:Brommer2007a}
to perform this optimisation. This implementation could then be used
to determine polarizable oxide potential parameters also for other
materials like alumina or magnesia.\\

\begin{acknowledgments}
  The authors thank P. Tangney for providing his simulation
  program as a reference. Support from the DFG through Collaborative Research
  Centre 716, Project B.1 is gratefully acknowledged.
\end{acknowledgments}


\begin{thebibliography}{23}
\expandafter\ifx\csname natexlab\endcsname\relax\def\natexlab#1{#1}\fi
\expandafter\ifx\csname bibnamefont\endcsname\relax
  \def\bibnamefont#1{#1}\fi
\expandafter\ifx\csname bibfnamefont\endcsname\relax
  \def\bibfnamefont#1{#1}\fi
\expandafter\ifx\csname citenamefont\endcsname\relax
  \def\citenamefont#1{#1}\fi
\expandafter\ifx\csname url\endcsname\relax
  \def\url#1{\texttt{#1}}\fi
\expandafter\ifx\csname urlprefix\endcsname\relax\def\urlprefix{URL }\fi
\providecommand{\bibinfo}[2]{#2}
\providecommand{\eprint}[2][]{\url{#2}}

\bibitem[{\citenamefont{{Iler}}(1979)}]{itapdb:Iler1979}
\bibinfo{author}{\bibfnamefont{R.~K.} \bibnamefont{{Iler}}},
  \emph{\bibinfo{title}{The chemistry of silica}} (\bibinfo{publisher}{Wiley,
  New York}, \bibinfo{year}{1979}), ISBN \bibinfo{isbn}{0--471--02404--X}.

\bibitem[{\citenamefont{{Karki} et~al.}(2007)\citenamefont{{Karki},
  {Bhattarai}, and {Stixrude}}}]{itapdb:Karki2007}
\bibinfo{author}{\bibfnamefont{B.~B.} \bibnamefont{{Karki}}},
  \bibinfo{author}{\bibfnamefont{D.}~\bibnamefont{{Bhattarai}}},
  \bibnamefont{and}
  \bibinfo{author}{\bibfnamefont{L.}~\bibnamefont{{Stixrude}}},
  \href{http://dx.doi.org/10.1103/PhysRevB.76.104205}{\bibinfo{journal}{Phys.\ Rev.~B} \textbf{\bibinfo{volume}{76}},
  \bibinfo{pages}{104205} (\bibinfo{year}{2007})}.

\bibitem[{\citenamefont{van {Beest} et~al.}(1990)\citenamefont{van {Beest},
  {Kramer}, and van {Santen}}}]{itapdb:Beest1990}
\bibinfo{author}{\bibfnamefont{B.~W.~H.} \bibnamefont{van {Beest}}},
  \bibinfo{author}{\bibfnamefont{G.~J.} \bibnamefont{{Kramer}}},
  \bibnamefont{and} \bibinfo{author}{\bibfnamefont{R.~A.} \bibnamefont{van
  {Santen}}}, \href{http://dx.doi.org/10.1103/PhysRevLett.64.1955}{\bibinfo{journal}{Phys.\ Rev.\ Lett.}
  \textbf{\bibinfo{volume}{64}}, \bibinfo{pages}{1955} (\bibinfo{year}{1990})}.

\bibitem[{\citenamefont{{Wilson} et~al.}(1996)\citenamefont{{Wilson}, {Madden},
  {Hemmati}, and {Angell}}}]{itapdb:Wilson1996}
\bibinfo{author}{\bibfnamefont{M.}~\bibnamefont{{Wilson}}},
  \bibinfo{author}{\bibfnamefont{P.~A.} \bibnamefont{{Madden}}},
  \bibinfo{author}{\bibfnamefont{M.}~\bibnamefont{{Hemmati}}},
  \bibnamefont{and} \bibinfo{author}{\bibfnamefont{C.~A.}
  \bibnamefont{{Angell}}}, \href{http://dx.doi.org/10.1103/PhysRevLett.77.4023}{\bibinfo{journal}{Phys.\ Rev.\ Lett.}
  \textbf{\bibinfo{volume}{77}}, \bibinfo{pages}{4023} (\bibinfo{year}{1996})}.

\bibitem[{\citenamefont{{Wilson} and {Walsh}}(2000)}]{itapdb:Wilson2000}
\bibinfo{author}{\bibfnamefont{M.}~\bibnamefont{{Wilson}}} \bibnamefont{and}
  \bibinfo{author}{\bibfnamefont{T.~R.} \bibnamefont{{Walsh}}},
  \bibinfo{journal}{J.~Chem.\ Phys.} \href{http://dx.doi.org/10.1063/1.1320056}{ \textbf{\bibinfo{volume}{113}},
  \bibinfo{pages}{9180} (\bibinfo{year}{2000})}.

\bibitem[{\citenamefont{{Tangney} and {Scandolo}}(2002)}]{itapdb:Tangney2002}
\bibinfo{author}{\bibfnamefont{P.}~\bibnamefont{{Tangney}}} \bibnamefont{and}
  \bibinfo{author}{\bibfnamefont{S.}~\bibnamefont{{Scandolo}}},
  \href{http://dx.doi.org/10.1063/1.1513312}{\bibinfo{journal}{J.~Chem.\ Phys.} \textbf{\bibinfo{volume}{117}},
  \bibinfo{pages}{8898} (\bibinfo{year}{2002})}.

\bibitem[{\citenamefont{{Rowley} et~al.}(1998)\citenamefont{{Rowley}, {Jemmer},
  {Wilson}, and {Madden}}}]{itapdb:Rowley1998}
\bibinfo{author}{\bibfnamefont{A.~J.} \bibnamefont{{Rowley}}},
  \bibinfo{author}{\bibfnamefont{P.}~\bibnamefont{{Jemmer}}},
  \bibinfo{author}{\bibfnamefont{M.}~\bibnamefont{{Wilson}}}, \bibnamefont{and}
  \bibinfo{author}{\bibfnamefont{P.~A.} \bibnamefont{{Madden}}},
  \href{http://dx.doi.org/10.1063/1.476481}{\bibinfo{journal}{J.~Chem.\ Phys.} \textbf{\bibinfo{volume}{108}},
  \bibinfo{pages}{10209} (\bibinfo{year}{1998})}.

\bibitem[{\citenamefont{{Herzbach} et~al.}(2005)\citenamefont{{Herzbach},
  {Binder}, and {Muser}}}]{itapdb:Herzbach2005}
\bibinfo{author}{\bibfnamefont{D.}~\bibnamefont{{Herzbach}}},
  \bibinfo{author}{\bibfnamefont{K.}~\bibnamefont{{Binder}}}, \bibnamefont{and}
  \bibinfo{author}{\bibfnamefont{M.~H.} \bibnamefont{{Muser}}},
  \href{http://dx.doi.org/10.1063/1.2038747}{\bibinfo{journal}{J.~Chem.\ Phys.} \textbf{\bibinfo{volume}{123}},
  \bibinfo{pages}{124711} (\bibinfo{year}{2005})}.

\bibitem[{\citenamefont{{Paramore} et~al.}(2008)\citenamefont{{Paramore},
  {Cheng}, and {Berne}}}]{itapdb:Paramore2008}
\bibinfo{author}{\bibfnamefont{S.}~\bibnamefont{{Paramore}}},
  \bibinfo{author}{\bibfnamefont{L.}~\bibnamefont{{Cheng}}}, \bibnamefont{and}
  \bibinfo{author}{\bibfnamefont{B.~J.} \bibnamefont{{Berne}}},
  \href{http://dx.doi.org/10.1021/ct800244q}{\bibinfo{journal}{J.~Chem.\ Theory Comput.} \textbf{\bibinfo{volume}{4}},
  \bibinfo{pages}{1698} (\bibinfo{year}{2008})}.

\bibitem[{\citenamefont{{Madelung}}(1918)}]{itapdb:Madelung1918}
\bibinfo{author}{\bibfnamefont{E.}~\bibnamefont{{Madelung}}},
  \bibinfo{journal}{Phys.~Z.} \textbf{\bibinfo{volume}{19}},
  \bibinfo{pages}{524} (\bibinfo{year}{1918}).

\bibitem[{\citenamefont{{Ewald}}(1921)}]{itapdb:Ewald1921}
\bibinfo{author}{\bibfnamefont{P.~P.} \bibnamefont{{Ewald}}},
  \href{http://dx.doi.org/10.1002/andp.19213690304}{\bibinfo{journal}{Ann.\ Phys.\ (Leipzig)} \textbf{\bibinfo{volume}{64}},
  \bibinfo{pages}{253} (\bibinfo{year}{1921})}.

\bibitem[{\citenamefont{{Fincham}}(1994)}]{itapdb:Fincham1994}
\bibinfo{author}{\bibfnamefont{D.}~\bibnamefont{{Fincham}}},
 \href{http://dx.doi.org/10.1080/08927029408022180}{ \bibinfo{journal}{Mol.\ Sim.} \textbf{\bibinfo{volume}{13}},
  \bibinfo{pages}{1} (\bibinfo{year}{1994})}.

\bibitem[{\citenamefont{{Gibbon} and {Sutmann}}(2002)}]{itapdb:Gibbon2002}
\bibinfo{author}{\bibfnamefont{P.}~\bibnamefont{{Gibbon}}} \bibnamefont{and}
  \bibinfo{author}{\bibfnamefont{G.}~\bibnamefont{{Sutmann}}}, in
  \emph{\bibinfo{booktitle}{{Q}uantum {S}imulations of {C}omplex {M}any-{B}ody
  {S}ystems}}, edited by
  \bibinfo{editor}{\bibfnamefont{J.}~\bibnamefont{{Grotendorst}}},
  \bibinfo{editor}{\bibfnamefont{D.}~\bibnamefont{{Marx}}}, \bibnamefont{and}
  \bibinfo{editor}{\bibfnamefont{A.}~\bibnamefont{{Muramatsu}}}
  (\bibinfo{publisher}{John von Neumann Institute for Computing (NIC),
  J{\"u}lich}, \bibinfo{year}{2002}), vol.~\bibinfo{volume}{10}, pp.
  \bibinfo{pages}{467--506}.

\bibitem[{\citenamefont{{Wolf} et~al.}(1999)\citenamefont{{Wolf}, {Keblinski},
  {Phillpot}, and {Eggebrecht}}}]{itapdb:Wolf1999}
\bibinfo{author}{\bibfnamefont{D.}~\bibnamefont{{Wolf}}},
  \bibinfo{author}{\bibfnamefont{P.}~\bibnamefont{{Keblinski}}},
  \bibinfo{author}{\bibfnamefont{S.~R.} \bibnamefont{{Phillpot}}},
  \bibnamefont{and}
  \bibinfo{author}{\bibfnamefont{J.}~\bibnamefont{{Eggebrecht}}},
  \href{http://dx.doi.org/10.1063/1.478738}{\bibinfo{journal}{J.~Chem.\ Phys.} \textbf{\bibinfo{volume}{110}},
  \bibinfo{pages}{8254} (\bibinfo{year}{1999})}.

\bibitem[{\citenamefont{{Stadler} et~al.}(1997)\citenamefont{{Stadler},
  {Mikulla}, and {Trebin}}}]{itapdb:Stadler1997a}
\bibinfo{author}{\bibfnamefont{J.}~\bibnamefont{{Stadler}}},
  \bibinfo{author}{\bibfnamefont{R.}~\bibnamefont{{Mikulla}}},
  \bibnamefont{and} \bibinfo{author}{\bibfnamefont{H.-R.}
  \bibnamefont{{Trebin}}},\href{ http://dx.doi.org/10.1142/S0129183197000990}{\bibinfo{journal}{Int.\ J.\ Mod.\ Phys.~C}
  \textbf{\bibinfo{volume}{8}}, \bibinfo{pages}{1131} (\bibinfo{year}{1997})},\\
  \bibinfo{note}{{\tt \href{http://www.itap.physik.uni-stuttgart.de/~imd/}{http://www.itap.physik.uni-stuttgart.de/\~{}imd/}}}.

\bibitem[{\citenamefont{{Ercolessi} and {Adams}}(1994)}]{itapdb:Ercolessi1994}
\bibinfo{author}{\bibfnamefont{F.}~\bibnamefont{{Ercolessi}}} \bibnamefont{and}
  \bibinfo{author}{\bibfnamefont{J.~B.} \bibnamefont{{Adams}}},
  \href{http://dx.doi.org/10.1209/0295-5075/26/8/005}{\bibinfo{journal}{Europhys.\ Lett.} \textbf{\bibinfo{volume}{26}},
  \bibinfo{pages}{583} (\bibinfo{year}{1994})}.

\bibitem[{\citenamefont{{Gaetani} et~al.}(1998)\citenamefont{{Gaetani},
  {Asimow}, and {Stolper}}}]{itapdb:Gaetani1998}
\bibinfo{author}{\bibfnamefont{G.~A.} \bibnamefont{{Gaetani}}},
  \bibinfo{author}{\bibfnamefont{P.~D.} \bibnamefont{{Asimow}}},
  \bibnamefont{and} \bibinfo{author}{\bibfnamefont{E.~M.}
  \bibnamefont{{Stolper}}}, \href{http://dx.doi.org/10.1016/S0016-7037(98)00172-0}{\bibinfo{journal}{Geochim.\ Cosmochim.\ Acta}
  \textbf{\bibinfo{volume}{62}}, \bibinfo{pages}{2499} (\bibinfo{year}{1998})}.

\bibitem[{\citenamefont{{Barrat} et~al.}(1997)\citenamefont{{Barrat}, {Badro},
  and {Gillet}}}]{itapdb:Barrat1997}
\bibinfo{author}{\bibfnamefont{J.-L.} \bibnamefont{{Barrat}}},
  \bibinfo{author}{\bibfnamefont{J.}~\bibnamefont{{Badro}}}, \bibnamefont{and}
  \bibinfo{author}{\bibfnamefont{P.}~\bibnamefont{{Gillet}}},
 \href{http://dx.doi.org/10.1080/08927029708024165}{\bibinfo{journal}{Mol.\ Sim.} \textbf{\bibinfo{volume}{20}},
  \bibinfo{pages}{17} (\bibinfo{year}{1997})}.

\bibitem[{\citenamefont{{Levien} et~al.}(1980)\citenamefont{{Levien},
  {Prewitt}, and {Weinder}}}]{itapdb:Levien1980}
\bibinfo{author}{\bibfnamefont{L.}~\bibnamefont{{Levien}}},
  \bibinfo{author}{\bibfnamefont{C.~T.} \bibnamefont{{Prewitt}}},
  \bibnamefont{and} \bibinfo{author}{\bibfnamefont{D.~J.}
  \bibnamefont{{Weinder}}}, \bibinfo{journal}{Am.\ Mineral.}
  \textbf{\bibinfo{volume}{65}}, \bibinfo{pages}{920} (\bibinfo{year}{1980}).

\bibitem[{\citenamefont{{Schmahl} et~al.}(1992)\citenamefont{{Schmahl},
  {Swainson}, {Dove}, and {Graeme-Barber}}}]{itapdb:Schmahl1992}
\bibinfo{author}{\bibfnamefont{W.~W.} \bibnamefont{{Schmahl}}},
  \bibinfo{author}{\bibfnamefont{I.~P.} \bibnamefont{{Swainson}}},
  \bibinfo{author}{\bibfnamefont{M.~T.} \bibnamefont{{Dove}}},
  \bibnamefont{and}
  \bibinfo{author}{\bibfnamefont{A.}~\bibnamefont{{Graeme-Barber}}},
  \bibinfo{journal}{Z.~Kristallogr.} \textbf{\bibinfo{volume}{201}},
  \bibinfo{pages}{125} (\bibinfo{year}{1992}).

\bibitem[{\citenamefont{{Levien} and {Prewitt}}(1981)}]{itapdb:Levien1981}
\bibinfo{author}{\bibfnamefont{L.}~\bibnamefont{{Levien}}} \bibnamefont{and}
  \bibinfo{author}{\bibfnamefont{C.~T.} \bibnamefont{{Prewitt}}},
  \bibinfo{journal}{Am.\ Mineral.} \textbf{\bibinfo{volume}{66}},
  \bibinfo{pages}{324} (\bibinfo{year}{1981}).

\bibitem[{\citenamefont{{Demontis} et~al.}(2001)\citenamefont{{Demontis},
  {Spanu}, and {Suffritti}}}]{itapdb:Demontis2001}
\bibinfo{author}{\bibfnamefont{P.}~\bibnamefont{{Demontis}}},
  \bibinfo{author}{\bibfnamefont{S.}~\bibnamefont{{Spanu}}}, \bibnamefont{and}
  \bibinfo{author}{\bibfnamefont{G.~B.} \bibnamefont{{Suffritti}}},
\href{ http://dx.doi.org/10.1063/1.1364638}{ \bibinfo{journal}{J.~Chem.\ Phys.} \textbf{\bibinfo{volume}{114}},
  \bibinfo{pages}{7980} (\bibinfo{year}{2001})}.

\bibitem[{\citenamefont{{Brommer} and
  {G{\"a}hler}}(2007)}]{itapdb:Brommer2007a}
\bibinfo{author}{\bibfnamefont{P.}~\bibnamefont{{Brommer}}} \bibnamefont{and}
  \bibinfo{author}{\bibfnamefont{F.}~\bibnamefont{{G{\"a}hler}}},
  \href{http://dx.doi.org/10.1088/0965-0393/15/3/008}{\bibinfo{journal}{Modelling Simul. Mater. Sci. Eng.}
  \textbf{\bibinfo{volume}{15}}, \bibinfo{pages}{295} (\bibinfo{year}{2007})},\\
  \bibinfo{note}{{\tt
  \href{http://www.itap.physik.uni-stuttgart.de/~imd/potfit/}{\footnotesize http://www.itap.physik.uni-stuttgart.de/\~{}imd/potfit/}}}.

\end{thebibliography}
\end{document}